\documentclass{elsart}

\usepackage{psfig}

\usepackage{natbib}

\def\ltsima{$\; \buildrel < \over \sim \;$}
\def\simlt{\lower.5ex\hbox{\ltsima}}
\def\gtsima{$\; \buildrel > \over \sim \;$}
\def\simgt{\lower.5ex\hbox{\gtsima}}

\begin{document}

\runauthor{M.Guainazzi et al.}


\begin{frontmatter}

\title{An exploratory study of the hard X-ray variability properties of PG quasars with RXTE}

\author[MGU]{M.\ Guainazzi}
\address[MGU]{XMM-SOC, VILSPA, ESA, Apartado 50727, E-28080 Madrid, Spain}
\author[FFI]{F.\ Fiore, E.\ Giallongo}
\address[FFI]{Osservatorio Astronomico di Roma, Via dell'Osservatorio 5, I-00040 Monteporzio Catone, Italy}
\author[ALA]{A.\ Laor}
\address[ALA]{Department of Physics, Technion, Israel Institute of Technology, Haifa 32000, Israel}
\author[MEL]{M.\ Elvis, A.\ Siemiginowska}
\address[MEL]{Harvard-Smithsonian Center for Astrophysics, 60 Garden st., Cambridge Ma., 02138, USA}

\begin{abstract}
We have monitored with the RXTE PCA
the variability pattern of the 2--20~keV flux
in four PG quasars (QSOs)
from the Laor et al. (1994) sample.  Six observations of each target
at regular intervals of 1 day were performed.
The sample comprises objects with extreme values of Balmer line
width (and hence soft X-ray steepness) and spans about one
order of magnitude in luminosity.
The most robust result is that
the variability amplitude decreases as energy increases.
Several options for a possible ultimate driver of the soft
and hard X-ray variability, such as the influx rate
of Comptonizing relativistic particles, instabilities in the accretion flow
or the number of X-ray ``active sites'', are consistent with our results. 
\end{abstract}


\end{frontmatter}


The
study that is the subject of this paper 
aims at a systematic investigation
of the variability properties
of a sizable sample of optically selected QSOs (the first after the
seminal paper by Zamorani et al 1984).
The sample is the Laor et al. (1997) sample of 23 PG quasars, selected
to have ${\rm z < 0.4}$, Galactic ${\rm N_H < 1.9 \times 10^{20}}$~cm$^{-2}$
and ${\rm M_B < 23}$.
The timing results of a monitoring campaign
of six of them with the ROSAT/HRI are discussed by Fiore et al.
1998 (F98). They discovered large amplitude (factor of 2) and rapid
(timescale $\sim$1~day) variability.
``Steep'' QSOs ({\it i.e.}, those whose ROSAT/PSPC energy spectral index
${\rm \alpha_{sx}}$ is $\simgt 3$)
show systematically larger amplitude variation than
``flat'' ones. Laor et al. (1994) discovered a strong correlation between
${\rm \alpha_{sx}}$ and the Full Width Half Maximum
(FWHM) of the H${\beta}$ optical
lines. This correlation can
be explained if the size of the Broad-Line Region (BLR)
is uniquely determined
by the luminosity of the active nucleus and the BLR gas is virialized.
In this case, ${\rm L/L_{Edd}}$ scales inversely as the square of the bulk
velocity. In this framework, objects with steep ${\rm \alpha_{sx}}$ display
larger variability, implying a lower mass black hole and therefore a
higher ${\rm L/L_{Edd}}$.

An exploratory program to study the hard X-ray
variability properties of a sub-sample
of PG QSOs on timescales $\simgt$1~day
was started with the Rossi X-ray Timing Explorer (RXTE)  Observatory.
The objects are listed in
Tab.~\ref{tab1}.
\begin{table*}
\caption{Some properties of the observed sample.}
\begin{footnotesize}
\begin{tabular}{lccccccc} \hline \hline
PG source & z & ${\rm \alpha_{PSPC}}$  & FWHM H$_{\beta}$ & ${\rm N_{H_{Gal}}}$$^a$ & Exposure & PCA count rate & $L_X$$^b$ \\ 
& & & (km~s$^{-1}$) & ($10^{20}$~cm$^{-2})$  & (ks) & (s$^{-1}$) & ($10^{44}$~erg~s$^{-1}$)~ \\ \hline
0052+251 & 0.154 & $-1.4$ & 5200 & 4.8 & 40.8 & $4.35 \pm 0.07$$^c$ & $9.1 \pm^{1.4}_{1.2}$ \\
1202+281 & 0.165 & $-1.2$ & 5050 & 1.7 & 39.5 & $2.48 \pm 0.06$$^d$ & $9 \pm^3_2$ \\
1211+143 & 0.085 & $-2.0$ & 1860 & 2.7 & 48.4 & $1.84 \pm 0.04$$^d$ & $1.5 \pm^{0.4}_{0.3}$ \\
1440+356 & 0.077 & $-2.1$ &  1450 & 1.0 & 50.4 & $1.01 \pm 0.03$$^e$ & $0.9 \pm^{0.5}_{0.3}$ \\ \hline \hline
\end{tabular}

\noindent
$^a$after Dickey \& Lockman (1990)

\noindent
$^b$ in the 2--10~keV energy band

\noindent
$^c$in the 2--20~keV energy band

\noindent
$^d$in the 2--12~keV energy band

\noindent
$^e$in the 2--7~keV energy band

\label{tab1}
\end{footnotesize}
\end{table*}
The monitoring campaign consisted of six pointings for each target,
with intervals of  approximately one day between each pointing.

\begin{figure*}
\begin{center}
\hbox{
\psfig{figure=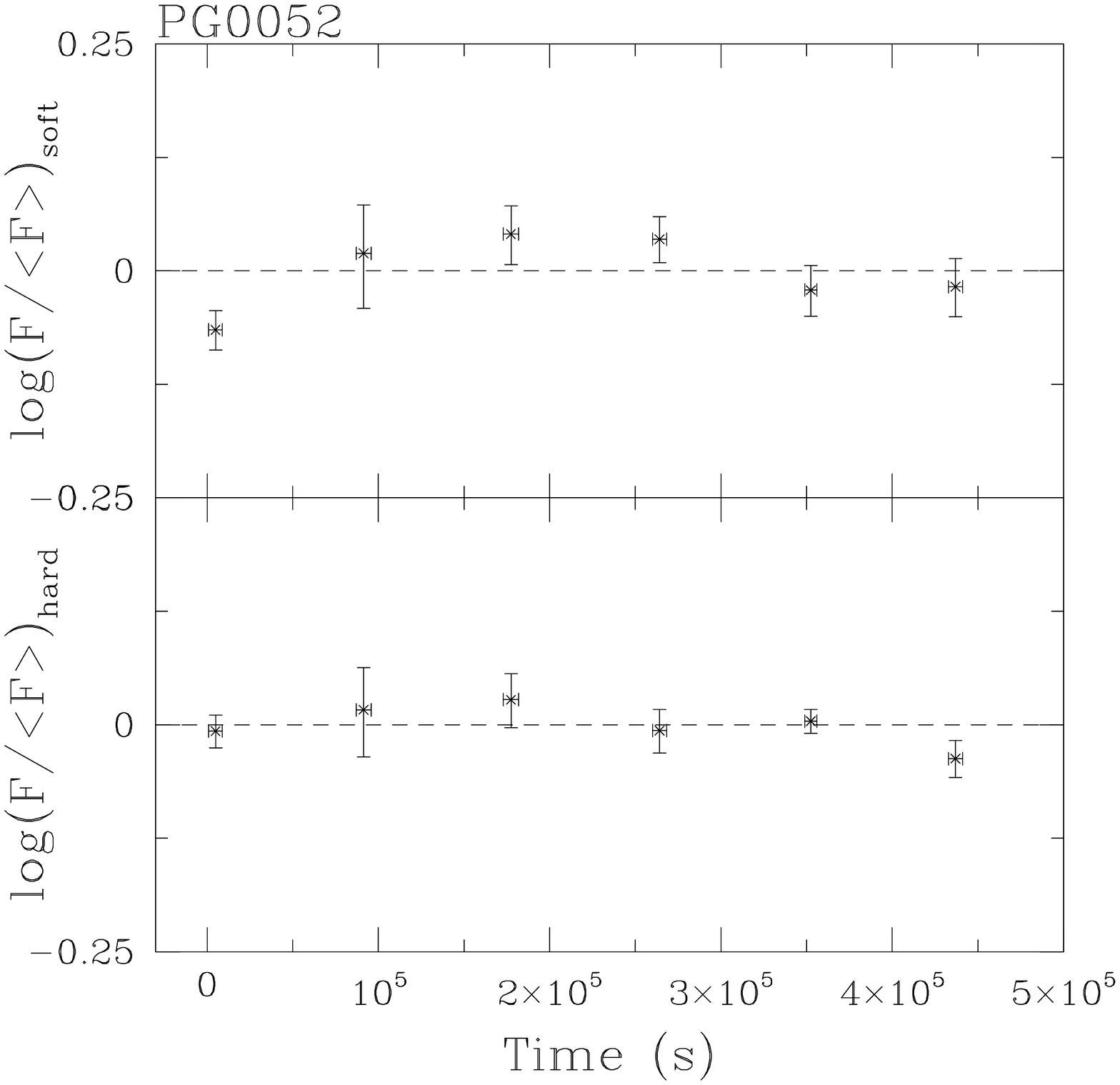,height=6.0cm,width=6.0cm,angle=0}
\hspace{1.0cm}
\psfig{figure=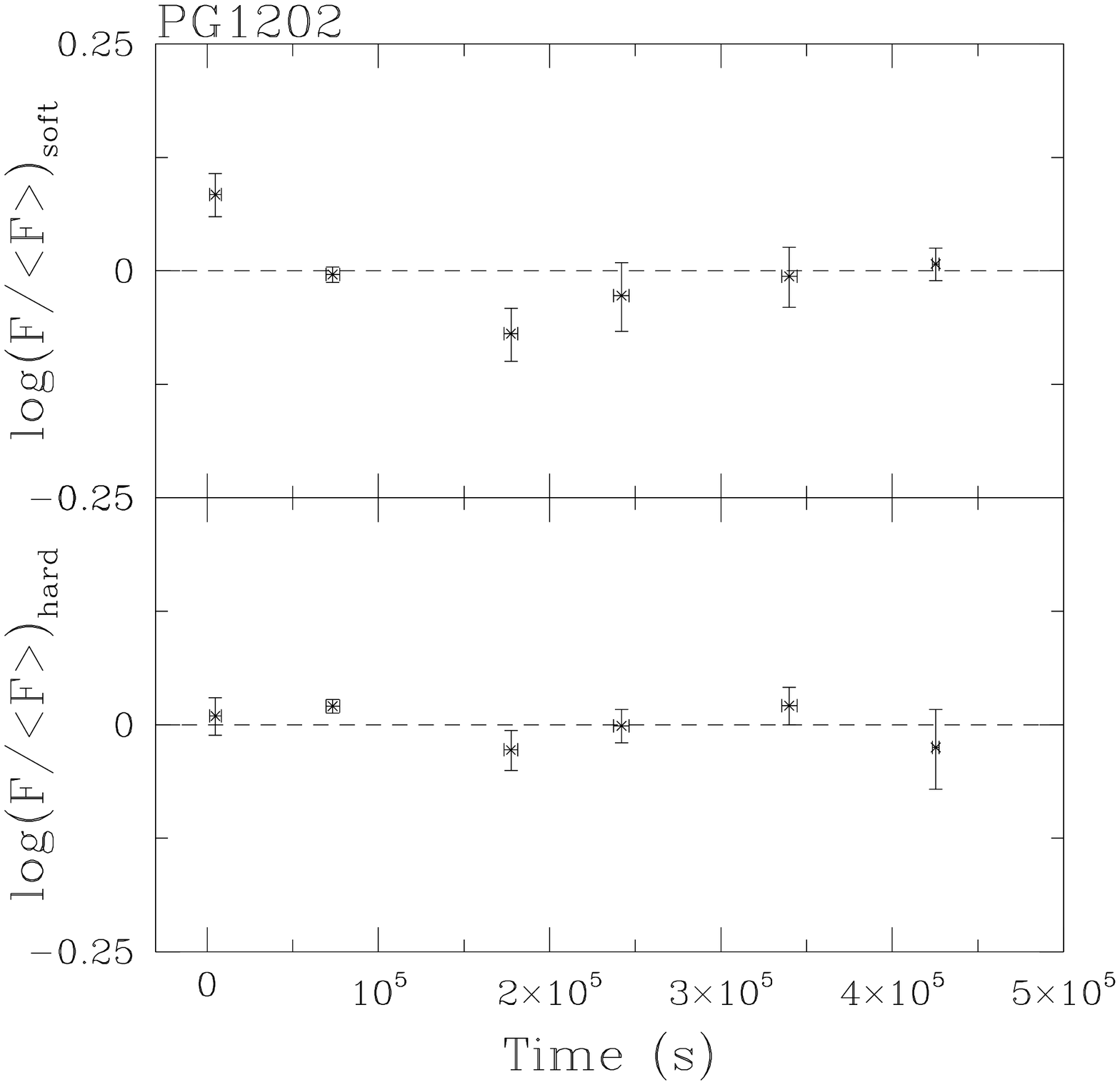,height=6.0cm,width=6.0cm,angle=0}
}
\hbox{
\psfig{figure=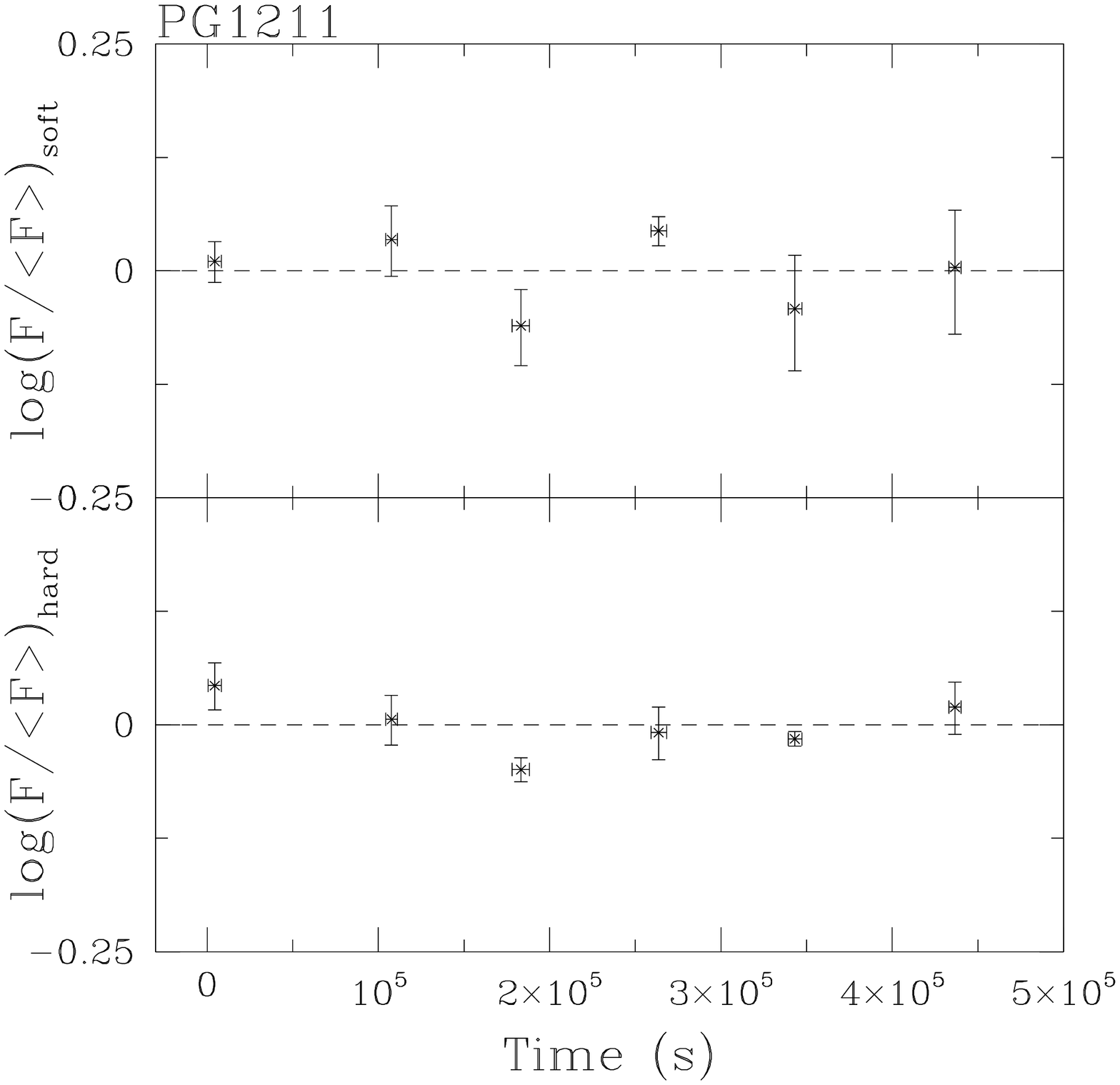,height=6.0cm,width=6.0cm,angle=0}
\hspace{1.0cm}
\psfig{figure=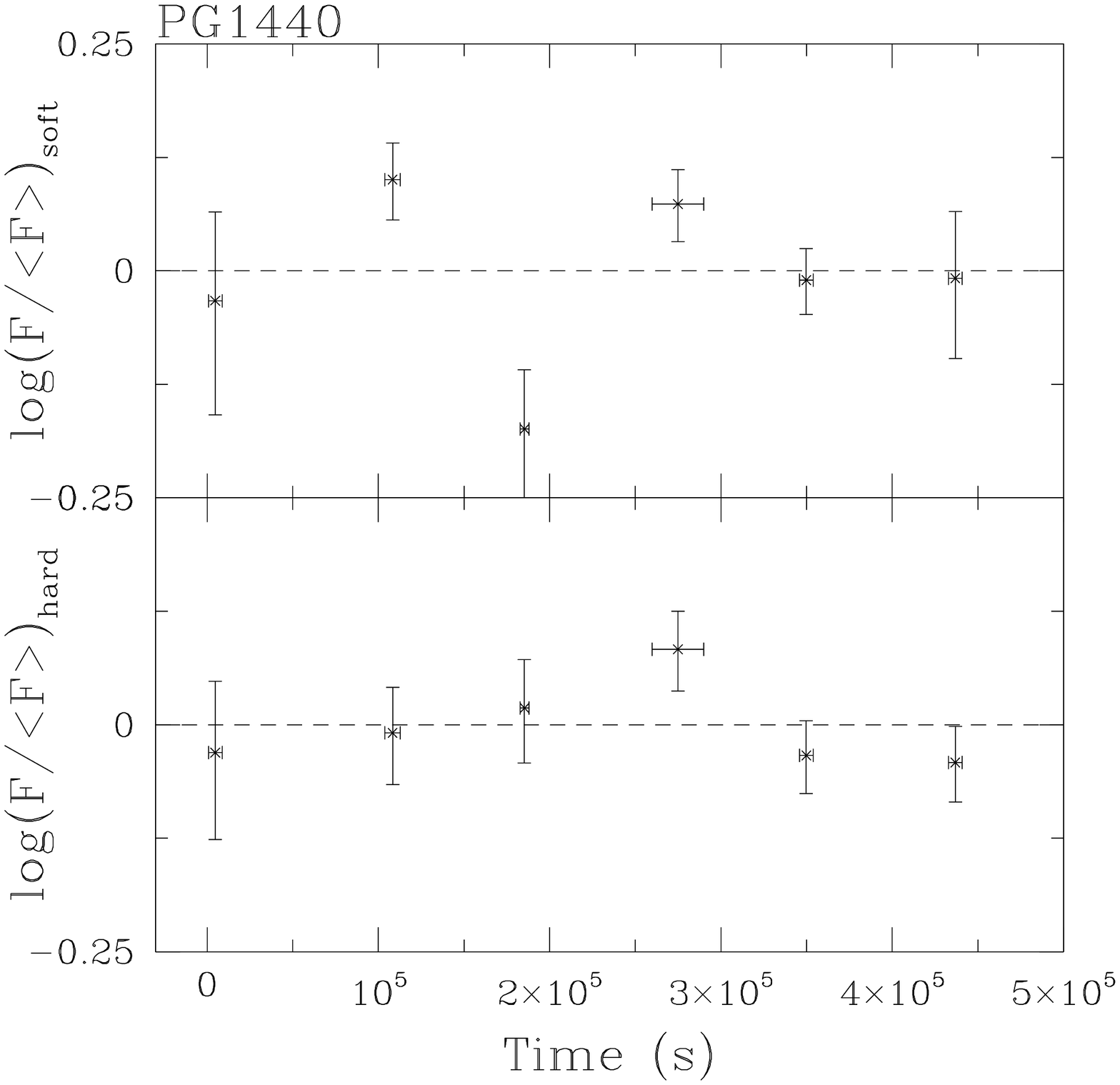,height=6.0cm,width=6.0cm,angle=0}
}
\end{center}
\vspace{-0.5cm}
\caption{Light curves (plotted as logarithm of the ratio of the flux to the
average flux) below 5~keV (``soft band'') and above 5~keV
(``hard band'').
The soft band is 2--5~keV. The hard bands are defined as in Tab.~\ref{tab1}}
\label{fig5}
\end{figure*}

\begin{figure}
\begin{center}
\hbox{
\psfig{figure=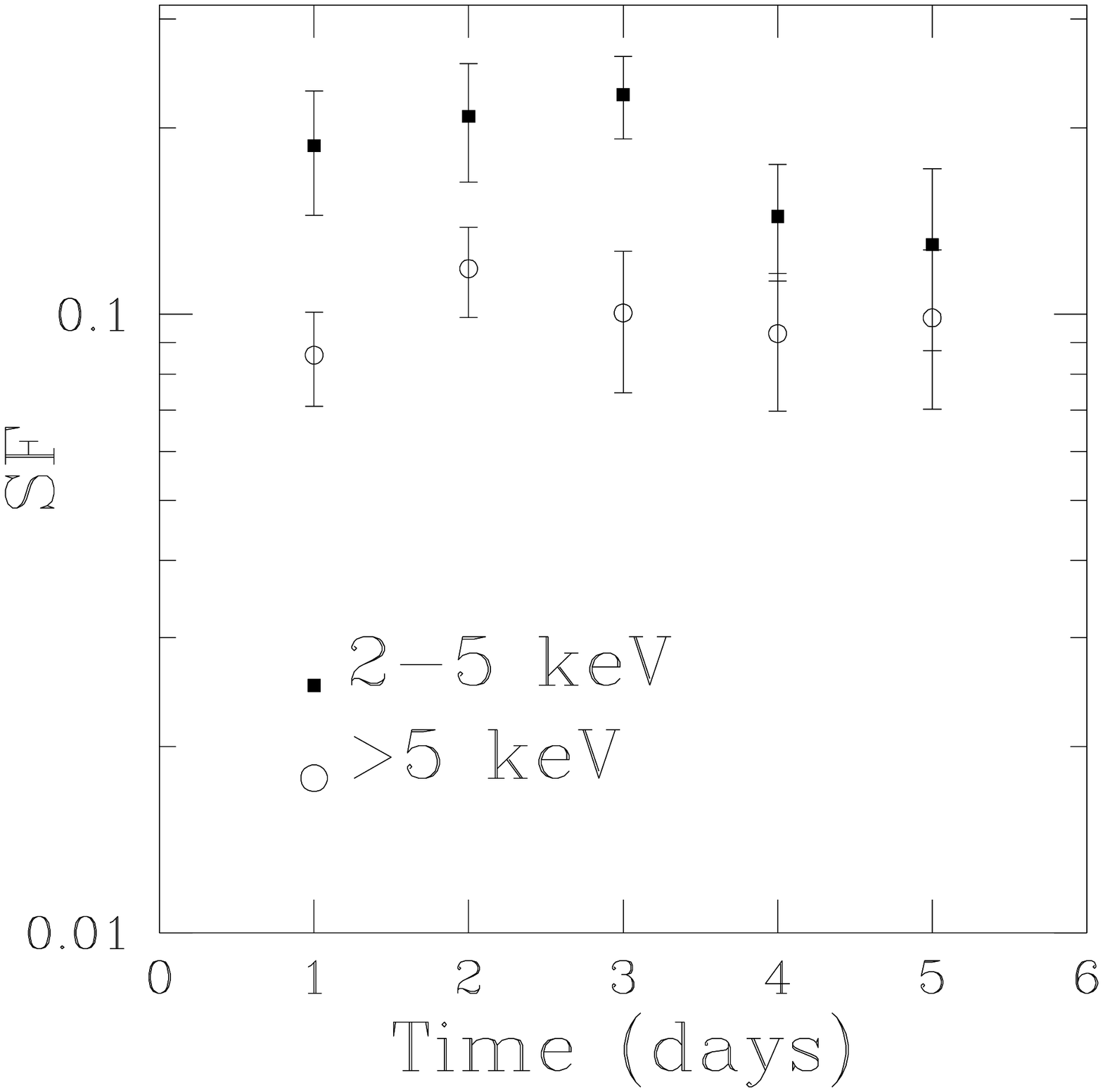,height=6.0cm,width=6.0cm,angle=0}
\hspace{0.5cm}
\psfig{figure=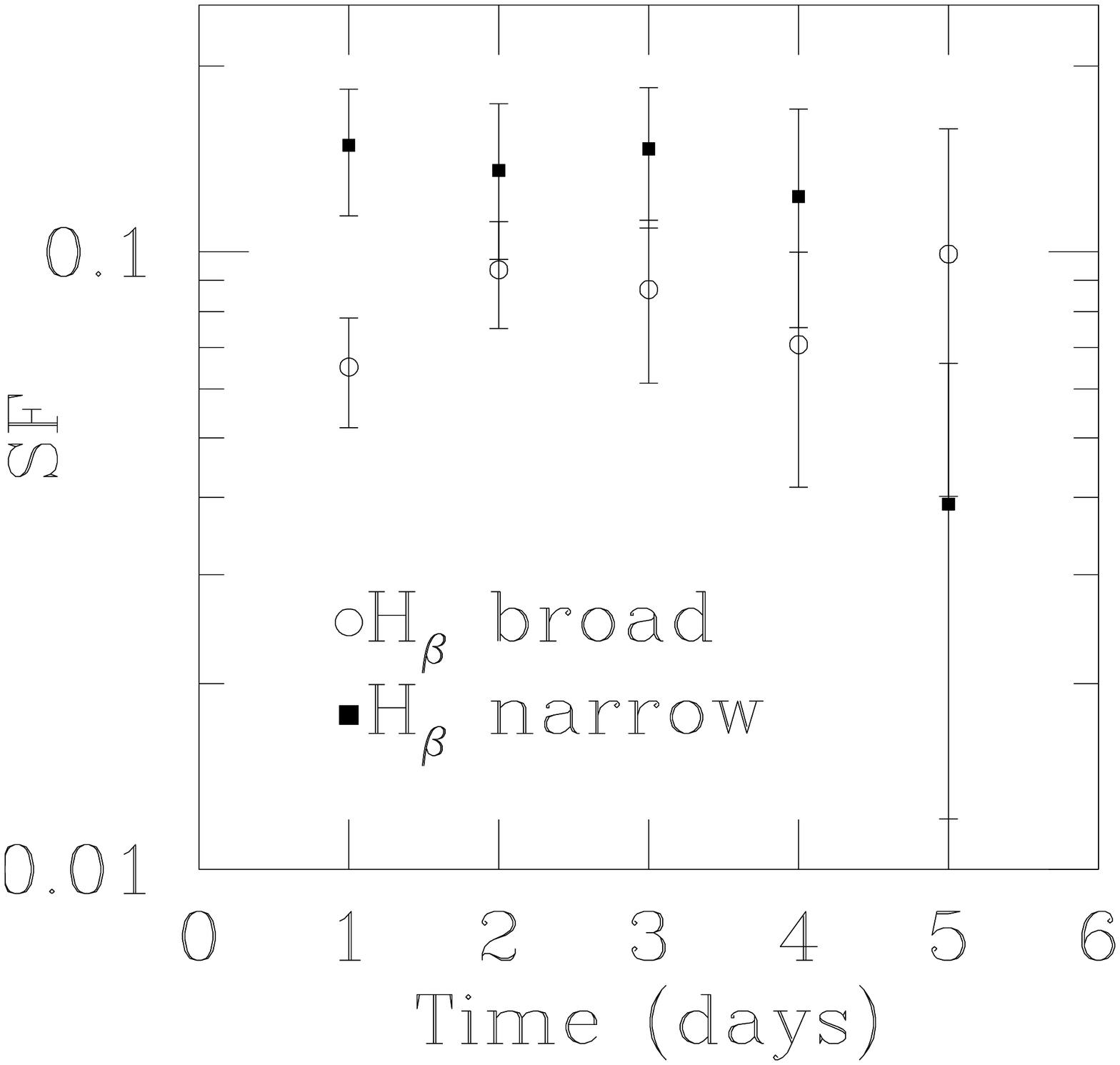,height=6.0cm,width=6.0cm,angle=0}
}
\end{center}
\vspace{-0.5cm}
\caption{{\it Left}: Structure function in the ``soft'' ({\it filled squares})
and ``hard'' ({\it open circles}) energy bands; {\it Right}: SF for quasars
with ``narrow'' ({\it i.e.}:
H$_{\beta}$ FWHM $<$2000~km~s$^{-1}$; {\it filled squares}) and
``broad'' ({\it open circles}) optical lines}
\label{fig9}
\end{figure}

The 1-day averaged light curves for all sources of the sample
are shown in Fig.~\ref{fig5}. All light curves in Fig.~\ref{fig5}
exhibit some degree of
variability, and it is generally more marked in the soft than in the
hard energy band.
To characterize quantitatively the variability in our sample, we used
the  so-called {\it average structure function} (SF, Di Clemente et al.
1996). In the formulation suggested by F98, it consists of the
mean of the logarithmic
ratio between each pair of flux measurements:
$$
SF = \ < { |2.5 \times \log [f(t_j)]/[f(t_i)]|} >
$$
The errors on the SF
are given by the standard uncertainties on the average. For 
our light curves, sampled six times at regular intervals of approximately
1 day between consecutive observations, we can build a five-point SF.
We have calculated the SF in the soft and hard
energy bands for all the QSOs in our sample
(the thresholds between the bands are defined in Tab.~\ref{tab1}).
The ``soft'' SF is systematically
higher than the ``hard'' one on all timescales
(see the left panel of Fig.~\ref{fig9}), although the difference is
significant at more than the 1-$\sigma$ level only for  data points
corresponding to ${\rm \Delta t \equiv t_j - t_i} \le 3$~days.
The results are far less clear if objects with ``narrow'' and
``broad'' optical lines are compared (see the right panel in
Fig.~\ref{fig9}). However, the suggestion exists that narrow-line
QSOs remain more variable in the hard X-ray band (see Tab.~\ref{tab10}).
\begin{table*}
\caption{Average SF for ${\rm \Delta t \le 3}$~days, calculated from subsets
of our sample selected according to the width of the H$_{\beta}$ line.}
\begin{center}
\begin{tabular}{lcc} \hline \hline
& ${\rm E \le 5}$~keV & ${\rm E > 5}$~keV \\ \hline
\multicolumn{3}{l}{For H$_{\beta}$ \dots} \\
broad & $0.15 \pm 0.02$ & $0.070 \pm 0.007$ \\
narrow & $0.26 \pm 0.04$ & $0.15 \pm 0.02$ \\ \hline
\end{tabular}
\end{center}
\label{tab10}
\end{table*}

The cause of the lower variability amplitude in harder X-rays can be
multifold: a) Compton-reprocessing of the primary nuclear continuum
(for which, however, there is no convincing evidence yet; George et al.
2000); b) smearing induced by Comptonization on a variability pattern,
driven by instabilities in the accretion disk; c) smearing induced by the
larger number of interactions that higher energy photons undergo in the
Comptonizing plasma; d) the number of ``active sites'', whereby the
soft X-rays might be produced by the superposition of a large number of
``flares'' (originating, {\it e.g.} in an accretion disk corona), while
the harder photons are produced in a more homogeneous and probably
compact innermost region.


\end{document}